\documentstyle[11pt,fleqn]{article}
\catcode`\@=11
\@addtoreset{equation}{section}
\def\theequation{\arabic{section}.\arabic{equation}}
\def\appendix{\renewcommand{\thesection}{\Alph{section}}\setcounter{section}{0}
              \renewcommand{\theequation}
            {\mbox{\Alph{section}.\arabic{equation}}}\setcounter{equation}{0}}
\def\maketitle{\thispagestyle{empty}\setcounter{page}0\newpage
                \renewcommand{\thefootnote}{\arabic{footnote}}
                  \setcounter{footnote}0}
\renewcommand{\thanks}[1]{\renewcommand{\thefootnote}{\fnsymbol{footnote}}
               \footnote{#1}\renewcommand{\thefootnote}{\arabic{footnote}}}
\newcommand{\preprint}[1]{\hfill{\sl preprint - #1}\par\bigskip\par\rm}
\renewcommand{\title}[1]{\begin{center}\Large\bf #1\end{center}\rm\par\bigskip}
\renewcommand{\author}[1]{\begin{center}\Large #1\end{center}}
\newcommand{\address}[1]{\begin{center}\large #1\end{center}}

\def\dinfn{\smallskip Dipartimento di Fisica, Universit\`a di Trento\\
                           and Istituto Nazionale di Fisica Nucleare,\\
                                   Gruppo Collegato di Trento, Italia}

\def\Idinfn{\address{\dinfn}}
\newcommand{\email}[1]{e-mail: \sl #1@science.unitn.it\rm}
\newcommand{\femail}[1]{\thanks{\email{#1}}}
\newcommand{\pacs}[1]{\smallskip\noindent{\sl PACS numbers:
                       \hspace{0.3cm}#1}\par\bigskip\rm}
\def\babs{\hrule\par\begin{description}\item{Abstract: }\it}
\def\eabs{\par\end{description}\hrule\par\medskip\rm}
\renewcommand{\date}[1]{\par\bigskip\par\sl\hfill #1\par\medskip\par\rm}
\newcommand{\ack}[1]{\par\section*{Acknowledgments} #1}
\newcommand{\s}[1]{\section{#1}}

\renewcommand{\vec}[1]{{\bf #1}}       
\def\hs{\qquad}               
\def\nn{\nonumber}            
\def\beq{\begin{eqnarray}}    
\def\eeq{\end{eqnarray}}      
\def\ap{\left.}               
\def\at{\left(}               
\def\aq{\left[}               
\def\ag{\left\{}              
\def\cp{\right.}              
\def\ct{\right)}              
\def\cq{\right]}              
\def\cg{\right\}}             
\newtheorem{theorem}{Theorem}                  
\newtheorem{lemma}{Lemma}                      
\newtheorem{proposition}{Proposition}          
\def\R{{\hbox{{\rm I}\kern-.2em\hbox{\rm R}}}}   
\def\H{{\hbox{{\rm I}\kern-.2em\hbox{\rm H}}}}   
\def\N{{\hbox{{\rm I}\kern-.2em\hbox{\rm N}}}}   
\def\C{{\ \hbox{{\rm I}\kern-.6em\hbox{\bf C}}}} 
\def\Z{{\hbox{{\rm Z}\kern-.4em\hbox{\rm Z}}}}   
\def\ii{\infty}                                  
\def\X{\times\,}                                 
\newcommand{\fr}[2]{\mbox{$\frac{#1}{#2}$}}      
\def\Tr{\mathop{\rm Tr}\nolimits}                  
\renewcommand{\Re}{\mathop{\rm Re}\nolimits}       
\renewcommand{\Im}{\mathop{\rm Im}\nolimits}       
\def\lap{\Delta}                                   

\def\ga{\gamma}
\def\de{\delta}
\def\ep{\varepsilon}
\def\ze{\zeta}

\def\ka{\kappa}
\def\la{\lambda}

\def\si{\sigma}
\def\om{\omega}

\def\Ga{\Gamma}

\def\citen#1{%
\edef\@tempa{\@ignspaftercomma,#1, \@end, }
\edef\@tempa{\expandafter\@ignendcommas\@tempa\@end}%
\if@filesw \immediate \write \@auxout {\string \citation {\@tempa}}\fi
\@tempcntb\m@ne \let\@h@ld\relax \let\@citea\@empty
\@for \@citeb:=\@tempa\do {\@cmpresscites}%
\@h@ld}
\def\@ignspaftercomma#1, {\ifx\@end#1\@empty\else
   #1,\expandafter\@ignspaftercomma\fi}
\def\@ignendcommas,#1,\@end{#1}
\def\@cmpresscites{%
 \expandafter\let \expandafter\@B@citeB \csname b@\@citeb \endcsname
 \ifx\@B@citeB\relax 
    \@h@ld\@citea\@tempcntb\m@ne{\bf ?}%
    \@warning {Citation `\@citeb ' on page \thepage \space undefined}%
 \else
    \@tempcnta\@tempcntb \advance\@tempcnta\@ne
    \setbox\z@\hbox\bgroup 
    \ifnum\z@<0\@B@citeB \relax
       \egroup \@tempcntb\@B@citeB \relax
       \else \egroup \@tempcntb\m@ne \fi
    \ifnum\@tempcnta=\@tempcntb 
       \ifx\@h@ld\relax 
          \edef \@h@ld{\@citea\@B@citeB}%
       \else 
          \edef\@h@ld{\hbox{--}\penalty\@highpenalty \@B@citeB}%
       \fi
    \else   
       \@h@ld \@citea \@B@citeB \let\@h@ld\relax
 \fi\fi%
 \let\@citea\@citepunct
}
\def\@citepunct{,\penalty\@highpenalty\hskip.13em plus.1em minus.1em}%
\def\@citex[#1]#2{\@cite{\citen{#2}}{#1}}%
\def\@cite#1#2{\leavevmode\unskip
  \ifnum\lastpenalty=\z@ \penalty\@highpenalty \fi 
  \ [{\multiply\@highpenalty 3 #1
      \if@tempswa,\penalty\@highpenalty\ #2\fi 
    }]\spacefactor\@m}
\begin{document}

\preprint{UTF 382 -  hep-th/9608089}
\title{
Determinant of Laplacian on a non-compact 3-dimensional hyperbolic
manifold with finite  volume}
\author{Andrei A. Bytsenko\thanks{email: abyts@spin.hop.stu.neva.ru}}
\address{State Technical University, St. Petersburg 195251, Russia}
\author{Guido Cognola\femail{cognola} and
Sergio Zerbini\femail{zerbini}}
\Idinfn

\date{September 1996}

\babs
The functional determinant of Laplace-type operators on the
3-dimensional non-compact hyperbolic manifold
with invariant fundamental domain of finite volume is computed
by quadratures and making use of the related terms of the
Selberg trace formula.
\eabs


\s{Introduction}

In the last years there has been a lot of investigations about
functional determinants on  topologically nontrivial manifolds.
Most of them have been concerned with Riemann flat or spherical spaces
(see for example Refs.~\cite{camp90-196-1,eliz94b}
and references therein) or orbifold factors of
spheres \cite{kenn81-23-2884,chan93-395-407}.
The case of compact hyperbolic manifolds has also  been considered
(see for example
Refs.~\cite{dhok86-104-537,sarn87-110-113,voro87-110-439,dhok88-60-917,bolt90-130-581,byts93-8-2011,byts95-36-5084,eliz94b,byts96-266-1,byts96u-377}).
In this case one is dealing with 2-dimensional
$H^2/\Ga$ and 3-dimensional
$H^3/\Ga$ compact hyperbolic
manifolds, $H^N$ being the Lobachevsky space and $\Ga$  a
discrete group of isometries acting on  $H^N$ and containing
loxodromic, hyperbolic and elliptic elements (see
Refs.~\cite{venk90b,elst85-17-83,elst87-277-655,byts92-33-3108,byts96-266-1}).
Such manifolds are relevant in string theory and in cosmological scenarios.

For non-compact
Riemannian surfaces of finite area, the functional determinant of
Laplace operator has been computed in
Refs.~\cite{efra88-119-443,mull92-109-265}.
In this paper we extend the analysis to the 3-dimensional case,
considering a Laplace-type operator acting on functions in a
non-compact, 3-dimensional manifold $H^3/\Ga$. In our example,
the discrete subgroup of isometry can be chosen in the form
$SL(2,\Z+i\Z)/\{\pm Id\}$,
Id being an isolated identity element of $\Ga$.
It is generated by parabolic mappings
and is associated with a non-compact manifold having
an invariant fundamental domain of finite volume.

Making use of the Selberg trace formula, we shall investigate the
asymptotic expansion of the heat kernel trace
$\Tr\exp(-tL)$, $L$
being a Laplace-like operator. We shall find that
the presence of parabolic elements in $\Ga$ leads to the appearance of
logarithmic factor in the small $t$ asymptotic expansion. For non-compact
Riemannian surfaces of finite area, this fact has been observed in Refs.
\cite{efra88-119-443,mull92-109-265}.
In this case the meromorphic
continuation of the $\zeta$-function has been shown to be regular at
$s=0$, thus the determinant of the Laplacian has been evaluated by
means of the standard $\zeta$-function regularisation
\cite{ray71-7-145,hawk77-55-133}.
In the 3-dimensional case, we shall show that $\zeta(s|L)$
is still a meromorphic function regular at $s=0$,
allowing the use of $\zeta$-function regularisation.

In the computation of functional determinant of Laplacian
on the generalized cone, the
appearance of a non-standard logarithmic term
has been recently pointed out  in
Ref.~\cite{bord96u-89} and this fact was first noted by Cheeger
\cite{chee83-18-575} and others authors
\cite{call83-88-357,brun85-58-133}.
However, in Ref.~\cite{bord96u-89} the things
have been arranged in order to avoid the logarithmic term, by dealing
with a conformally coupled free massless field. A similar fact
has been recently shown to happen in an ultrastatic space-time of the form $\R \times H^3/\Ga$
\cite{byts96u-377}.

The contents of the paper are the following.
In Section~2 we summarize some properties of the combined contributions to the
Selberg trace formula we shall use in the paper.
In Section~3 the heat kernel trace and the $\zeta$-function for a
Laplace type operator are studied by making use of the trace formula.
In Section~4 the functional determinat is evaluated
by means of the quadrature method.
Finally we end with some conclusions in Section~5.

\s{Fundamental domain of the discrete group $SL(2,\Z+i\Z)/\{\pm Id\}$
and the Selberg trace formula associated with the
cusp form}

Here we summarise the geometry and local isometry
associated with a simple 3-dimensional complex Lie group. We shall
consider discrete subgroup $\Ga\subset SL(2,\C)/\{\pm Id\}$, where
$Id$ is the $2\X2$ identity matrix and is an isolated element of the
$\Ga$. The group $\Ga$ acts discontinuously at
the point $z\in\bar\C$, $\bar\C$ being the extended complex plane.
We recall that a transformation $\ga\neq Id$, $\ga\in\Ga$, with
\beq
\ga z=\frac{az+b}{cz+d}\,,\hs ad-bc=1\,,
\hs (\Tr\ga)^2=(a+d)^2\,, \hs a,b,c,d\in\C,
\eeq
is called elliptic if $(\Tr\ga)^2$ satisfies
$0\leq(\Tr\ga)^2<4$, hyperbolic if $(\Tr\ga)^2>4$,
parabolic if $(\Tr\ga)^2=4$ and loxodromic
if $(\Tr\ga)^2\in\C\backslash\aq0,4\cq$.
The classification of these
transformations can also be based on the properties of their fixed
points, the number of which is one for the parabolic transformations
and two for all other cases.

The element $\ga\in SL(2,\C)$ acts on $z=(y,w)\in H^3$,
$w=x_1+ix_2$ by means of the following
linear-fractional transformation:
\beq
\ga z=\at\frac{y}{|cw+d|^2+|c|^2y^2}\,,\frac{(aw+b)(\bar c \bar
w+\bar d)+a \bar c y^2}{|cw+d|^2+|c|^2y^2} \ct
\:.\eeq
The isometric circle of a transformation
$\ga\in SL(2,\C)/\{\pm Id\}$
for which $\ii$ is not a fixed point is defined to be the circle
\beq
I(\ga)=\{z:\:\:\:|\ga z|=1\}\:,\hs\mbox{ or}\hs
I(\ga)=\{z:\:\:\:|z+d/c|=|c|^{-1}\}\:,\:\:\: c\neq0\:.
\eeq
A transformation $\ga$ carries its isometric circle
$I(\ga)$ into $I(\ga^{-1})$.

The isometric fundamental domain of a Fuchsian group (Kleinian group
without loxodromic elements) has the following  structure: it is bounded by
arcs of circles orthogonal to the invariant circle and consists either
of two symmetric components or of a single component, while the
mappings connecting its equivalent sides, generate the whole group. In
many cases, it is more convenient to deal with other fundamental
regions. For example, the so-called normal fundamental Dirichlet's
polygons are often used for Fuchsian groups
and we shall follow this approach here.

Now we consider a discrete subgroup of a special kind.
Let $G=PSL(2,\C)=SL(2,\C)/\{\pm Id\}$, then for $\Ga\subset G$, one can choose
a subgroup of $\Ga$ in the form $SL(2,\Z+i\Z)/\{\pm Id\}$, where $\Z$ is the
ring of integer numbers. The element $\ga\in\Ga$ will be identified with
$-\ga$. The group $\Ga$ has, within a conjugation, one maximal
parabolic subgroup $\Ga_\ii$ ($c=0$). Thus, the fundamental domain
related to $\Ga$ has one parabolic vertex and can be taken in the form
\cite{venk73-125-3,lang85b}
\beq
F(\Ga)=\ag (y,w): x_1^2+ x_2^2 +y^2 >1,\:\:\:
-\fr12<x_2<x_1<\fr12\cg\:.\label{a2}\eeq
{\bf Remark}. {\it Let a free abelian group of isometries be generated by
the two parabolic mappings
\beq
g_1(z)=z+1\,,\hs  g_2(z)=z+i
\:,\eeq
then, if we identify the faces of the polyhedron, Eq.~(\ref{a2}), we get a
manifold $M(\Ga)$ that is homeomorphic to a punctured torus
$S^1\otimes S^1\otimes\aq-\frac{1}{2},\frac{1}{2}\ct=U_c\otimes S^1$,
where $U_c=\ag z:\:\:\:0<|z|\leq\frac{1}{2}\cg$ is a punctured cylinder.
It is turned into a hyperbolic manifold by removing the boundary
$\partial M(\Ga)$,
which is homeomorphic to the torus $S^1\otimes S^1$.}

Now we are ready to  start the discussion of the Selberg trace formula,
which can be constructed as an expansion in eigenfunctions of
the automorphic Laplacian. To begin with, we assume that the group
$\Ga$ is generated by parabolic mappings. Since the discrete group
$\Ga$ has a cusp at $\ii$ ($c=0$), each element of the stabiliser
$\Ga_{\ii}$ is a translation.
Computing the conjugacy class
$\ag\ga\cg_{\Ga}$, $\ga\in\Ga_{\ii}$
with $\ga$ different from identity, one easily gets
\begin{proposition}
Let
\beq
\ga=\left(\matrix{1& n_1+in_2\cr0&1\cr}\right)
\:,\hs n_1,n_2\in\Z \:.
\label{b1}\eeq
The conjugacy class with representative $\ga$ consists in element  $\ga$  and $\ga^{-1}$, where
\beq
\ga^{-1}=\left(\matrix{1&-n_1-in_2\cr0&1\cr} \right)
\:.\label{b2}
\eeq
The remaining conjugacy classes have the representatives in
$\Ga_{\ii}$ of the form
\beq
\ga_1=\left(\matrix{i&0\cr0& -i\cr}\right)\:,\,\,\,
\ga_2=\left(\matrix{i&1\cr0& -i\cr}\right)\:,\,\,\,
\ga_3=\left(\matrix{i&-i\cr0&-i\cr}\right)\:,\,\,\,
\ga_4=\left(\matrix{i&1-i\cr0&-i\cr}\right)\:.
\eeq
The centralisers related to these representations read
\beq
\Ga^\ga=\left(\matrix{1&m_1+im_2\cr0&1\cr} \right)
\:,\hs  m_1,m_2\in \Z\:,
\eeq
\beq
\Ga^1=\Ga^{\ga_1}=\ag
\left(\matrix{1&0\cr0&1\cr}\right)\:,
\left(\matrix{i&1\cr0&-i\cr}\right)\:,
\left(\matrix{0&1\cr -1&0\cr}\right)\:,
\left(\matrix{0&i\cr i&0\cr}\right)
\cg\:,\nn\eeq
\beq
\Ga^2=\Ga^{\ga_2}=\ag
\left(\matrix{1&0\cr0&1\cr}\right)\:,
\left(\matrix{i&1\cr0&-i\cr}\right)\:,
\left(\matrix{i&0\cr2& -i\cr}\right)\:,
\left(\matrix{-1&i\cr 2i&1\cr}\right)
\cg\:,\nn\eeq
\beq
\Ga^3=\Ga^{\ga_3}=\ag
\left(\matrix{1&0\cr0&1\cr}\right)\:,
\left(\matrix{i&-i\cr0&-i\cr}\right)\:,
\left(\matrix{1&-1\cr2&-1\cr}\right)\:,
\left(\matrix{i&0\cr2i&-i\cr}\right)
\cg\:,\nn\eeq
\beq
\Ga^4=\Ga^{\ga_4}=\ag
\left(\matrix{1&0\cr0&1\cr}\right)\:,
\left(\matrix{i&1-i\cr0&-i\cr}\right)\:,
\left(\matrix{i&0\cr1+i&-i\cr}\right)\:,
\left(\matrix{1&-1-i\cr1-i&-1\cr}\right)
\cg\:.\nn\eeq
\end{proposition}

Let us consider an arbitrary integral operator with kernel $k(z,z')$.
Invariance of the operator is equivalent to fulfillment of the condition
$k(\ga z,\ga z')=k(z,z')$ for any $z,z'\in H^3$ and
$\ga\in PSL(2,\C)$. So the kernel of the invariant operator is a
function of the geodesic distance between $z$ and $z'$.
It is convenient to replace such a distance with the fundamental
invariant of a pair of points
$u(z,z')=|z-z'|^2/yy'$, thus  $k(z,z')=k(u(z,z'))$ . Let $\la_i$ be
the isolated eigenvalues of the self-adjoint extension of the Laplace operator
and let us introduce a suitable analytic function $h(r)$ and
$r^2_j=\la_j-1$. It can be shown that $ h(r)$
is related to the quantity $k(u( z,\ga z))$
by means of the Selberg transform. Let us denote by $g(u)$ the
Fourier transform of $ h(r)$, namely
\beq
g(u)=\frac{1}{2\pi}\int_{-\ii}^\ii e^{-iru} h(r) dr
\:.\eeq

For one parabolic vertex  let us introduce a subdomain $F_Y$ of the fundamental
region $F(\Ga)$ by means
\beq
F_Y=\ag z \in F(\Ga), z=\ag y, \vec x \cg | y \leq Y \cg
\:,\label{b7}\eeq
where $ Y$ is a sufficiently large positive number.
\begin{lemma}\label{L1}
Suppose $h(r)$ to be an even analytic function in the strip
$|\Im r|<1+\ep$ ($\ep>0$) and  $h(r)=O(1+|r|^2)^{-2}$.
Then for $N=3$ the following formula holds \cite{venk73-125-3}:
\beq
\sum_j h(r_j)&=&\lim_{Y\to\ii}\ag
\int_{F_Y}\sum_{\{\ga\}_{\Ga}}k(u(z,\ga z))\:d\mu(z)
\cp \nn\\
&&\hs\hs-\ap
\frac{1}{2\pi}\int_0^\ii h(r)\int_{F_Y}|E(z,1+ir)|^2\:d\mu(z)dr
\cg\:,\label{b8}\eeq
where $d\mu(z)=y^{-3}dydx_1dx_2$ is the invariant measure on $H^3$
and $E(z,s)$ is the
Eisenstein-Maass series associated with one cusp, namely
\beq
E(z,s)=\sum_{\ga\in(\Ga/\Ga_\ii)} y^s(\ga z)\,,\hs x_2(z)=\Im z
\:.\label{a8}\eeq
\end{lemma}

The series (\ref{a8}) converges absolutely for $\Re s>1$ and uniformly in $z$
on compact subset of $H^3$. All poles of $E(z,s)$ are contained in
the union of the region $\Re s<1/2$ and the interval
$\aq1/2,1\cq$ and those contained in such an interval are simple.
Furthermore, for each $s$, the series  $E(z,s)$ is a real
analytic function on $H^3$, automorphic relative to the group $\Ga$
and satisfies the eigenvalues equation
\beq
\lap E(z,s)=s(s-1)E(z,s)
\:,\label{b9}\eeq
$\lap$ being the  Laplace operator.
The asymptotic expansion of the second integral in Eq.~(\ref{b8}) can be
found with the help of Maass-Selberg relation \cite{venk73-125-3}.
For $Y\to\ii$ one has
\beq
\frac{1}{2\pi}\int_0^\ii h(r)\int_{F_Y}|E(z,1+ir)|^2\:d\mu(z)dr
&=&g(0)\ln Y+\frac{h(0)}{4}S(1)
\nn\\&&
-\frac{1}{4\pi}\int_{-\ii}^\ii  h(r)
\frac{S'(1+ir)}{S(1+ir)}\:dr
+O(1)
\:.\label{b10}\eeq
The function $S(s)$ (in the general case it is the S-matrix) is given
by a generalised Dirichlet series, convergent for $\Re s>1$,
\beq
S(s)=\frac{\sqrt\pi\Ga(s-\frac{1}{2})}{\Ga(s)}
\sum_{c\neq0}\sum_{0\leq d<|c|} |c|^{-2s}
\:,\label{b11}\eeq
where the sums are taken over all pairs $c,d$ of the matrix
$\at\matrix{*&*\cr c&d\cr}\ct\subset\Ga_\ii\backslash\Ga/\Ga_\ii$.
Also the poles of the meromorphic function $S(s)$ are contained
in the region $\Re s<1/2$ and in the interval $\aq1/2,1\cq$.
The functions $E(z,s)$ and $S(s)$ can
be analytically extented on the whole complex
s-plane, where they satisfy the functional equations
\beq
S(s)S(1-s)=Id\:,\hs
\overline{S(s)}=S(\bar s)\:,\hs
E(z,s)&=&S(s)E(z,1-s)
\:.\label{b12}\eeq

It should be noted that the terms of the trace formula associated with
the elements $\ga$ and $\ga^{-1}$ coincide. Then the contribution to
the first integral in Eq.~(\ref{b8}), which comes from all conjugacy
classes of the $\ga$-type
($\ga\in\Ga^{\ga}$), for $Y\to\ii$ can be written as follows
\beq
\int_{F_Y}\sum_{\{\ga\}_{\Ga_{\infty}}}k(u(z,\ga z))\:d\mu(z)
&=&\at \ln Y+ C\ct g(0)+\frac{h(0)}{4}
\nn\\&&\hs
-\frac{1}{4\pi}\int_{-\ii}^\ii h(r)
\psi(1+\fr{ir}{2})\:dr+O(1)
\:,\label{b13}\eeq
were $\psi(s)$ is the logarithmic derivative
of the Euler $\Ga$-function and $ C$ a computable constant which
reads \cite{venk73-125-3}
\beq
 C&=&\frac{5\ln2}{16}-\frac\ga2+C_0\:,\nn\\
C_0&=&\lim_{N\to\ii}\:\:\frac1{4\pi}
\sum_{i=1}^{N}\aq|\xi^i|^{-2}-2\pi\ln\frac{|\xi^{i+1}|}{|\xi^i|}\cq
-\frac12\ln|\xi^1|
\:.\eeq
In the latter equation $\ga$ is the Euler-Mascheroni constant
and $\xi$ is a two-dimensional vector, such that
$\ga z=\{y,\om+\xi\}$, $\xi\neq0$, $|\xi^{i+1}|\geq|\xi^i|$.

For the derivation of the Selberg trace formula,  one has to consider the contributions
coming from the identity and the non-parabolic elements in $\Ga$, the normalised
Eisenstein-Maass series, Eq.~(\ref{a8})
and all  $\ga$-type conjugacy classes,
Eq.~(\ref{b13}).  The final result we state should be considered as an
explicit addition to Lemma~\ref{L1}.
\begin{theorem}\label{Tstf}
For the special discrete group $SL(2,\Z+i\Z)/\{\pm Id\}$ and $h(r)$ satisfying
the conditions of Lemma \ref{L1}, we have the Selberg's trace formula
\beq
\sum_j h(r_j)&-&\sum_{\scriptstyle\{\ga\}_{\Ga},\ga\not=Id,
\atop\scriptstyle\ga-non-parabolic}\int k(u(z,\ga z))\:d\mu(z) - \frac{1}{4\pi}\int_{-\ii}^\ii h(r)
\frac{S'(1+ir)}{S(1+ir)}\:dr+\frac{h(0)}{4}S(1)
\nn\\ &=&
V(F)\int_0^\ii\frac{r^2}{2\pi^2}\:h(r)\:dr
+ Cg(0)+\frac{h(0)}{4}
-\frac{1}{4\pi}\int_{-\ii}^\ii h(r)
\psi(1+\fr{ir}{2})\:dr
\:.\label{stf}\eeq
\end{theorem}
The first term in the r.h.s. of Eq.~(\ref{stf}) is the contribution of
the identity element, while $V(F)$ is the (finite) volume
of the fundamental domain $F$ with respect to the measure $d\mu$.

\s{The heat kernel and the $\zeta$-function} \label{S:HK}

As discussed in the Introduction,
the determinant of an elliptic differential operator
requires a regularisation. It is convenient
to introduce the operator $L_\de=-\lap+\de^2-1$, with $\de$ a suitable
parameter. One of the most
used regularisation is the $\zeta$-function regularisation
\cite{ray71-7-145,dowk76-13-3224,hawk77-55-133}. By means of it one has
\beq
\ln\det L_\de=-\ze'\at 0|L_\de \ct
\:,\label{haw}\eeq
where the $\ze'$ is the derivative with respect to $s$ of
the $\zeta$-function. In the standard cases,
the $\zeta$-function at $s=0$ is well defined and so by means of the
latter formula one gets a finite result.

The meromorphic structure of the analytically continued
$\zeta$-function, as well as the ultraviolet divergences of the one-loop
effective action, can be related to the asymptotic properties
of the heat-kernel trace.
For the rank-1 symmetric space $H^3/\Ga$
the trace of the operator $\exp{[-(tL_\de)]}$ may be computed by using
Theorem \ref{Tstf} (Eq.~(\ref{stf})) with the choice
$h(r)=\exp\aq-t(r^2+\de^2)\cq$
(we use units in which the curvature $\ka=R/6$ of $H^3$ is equal to $-1$).
We have
\beq
g(u)=\frac{e^{-t \de^2}e^{-u^2/4t}}{\sqrt{4\pi t}}\,,\hs
g(0)=\frac{e^{-t \de^2}}{\sqrt{4\pi t}}\,,\hs
h(0)=e^{-t \de^2}\,.
\label{g}\eeq
In this and next Sections we shall consider additive terms of the
$\zeta$-function associated with identity and parabolic elements of group
$\Ga$ only (the heat kernel and $\zeta$-function analysis for co-compact
discrete group $\Ga$ has been done, for example, in Refs. [2,12]).

As a result
\beq
\Tr e^{-t L_\de}=e^{-t \de^2}\aq
\frac{V(F)}{(4\pi t)^{3/2}}
+\frac{ C}{(4\pi t)^{1/2}}
+\frac1{4}-\frac{1}{4\pi}\int_{-\ii}^{\ii}
\psi(1+\fr{ir}{2})e^{-tr^2}\:dr \cq\:.
\label{Kstf}
\eeq
The asymptotic behaviour of the last integral for $t\to0$
can be easily evaluated. In fact we may rewrite
\beq
\Tr e^{-t L_\de}=e^{-t \de^2}\aq \frac{\ln t}{8\sqrt{\pi t}}
+\frac{V(F)}{(4\pi t)^{3/2}}
+\frac{C+\ln2+\frac\ga4}{(4\pi t)^{1/2}}
-\frac{t}{\pi i}\int_{-\ii}^\ii  e^{-t r^2}f(\fr{ir}{2})\:dr \cq
\:.\label{stff}\eeq
The function $f(z)$ is defined by (see Ref.~\cite{grad80b})
\beq
f(z)=\frac12\sum_{k=1}^{\ii}\frac{k}{(k+1)(k+2)}
\sum_{n=1}^{\ii}\frac1{(n+z)^{k+1}}\,,
\label{fdiz}\eeq
and has an aymptotic expansion for large $|z|$ in terms of
the Bernoulli numbers $B_k$ given by
\beq
f(z)\sim\sum_{k=1}^{\ii}\frac{B_{2k}}{2k(2k-1)z^{2k-1}}
\:.\label{fdizExp}\eeq
The contribution for short $t$ comes from this asymptotics.
Thus we have
\begin{proposition}
The asymptotic behavior of the heat kernel for $t\to0$ reads
\beq
\Tr e^{-tL_\de}\simeq e^{-t\de^2}
\aq\frac{\ln t}{8\sqrt{\pi t}}+
\sum_{n=0}^\ii K_n t^{n-\frac32}\cq
=\sum_{r=0}^\ii
(A_r+P_r\ln t)t^{r-\frac32}
\:,\label{l3}\eeq
where the first $K_n$ coefficients are given by
\beq
K_0=\frac{V(F)}{(4\pi)^{3/2}}\,,\hs
K_1=\frac{C+\ln2+\frac\ga4}{\sqrt{4\pi}}\,,\hs
K_2=\frac{1}{6\sqrt{\pi}}\,,
\eeq
and
\beq
A_r=\sum_{n=0}^{r}(-1)^n\frac{B_{r-n}\de^{2n}}{n!}\:,\hs
P_0=0\:,\hs P_r=(-1)^{r-1}\frac{\de^{2(r-1)}}{8\sqrt{\pi}(r-1)!}
\:.\label{KP}\eeq
\end{proposition}
It should be noted that, besides the usual terms one has for the heat
kernel in 3-dimensions, there exist terms with logarithmic factors
due to the presence of parabolic elements in $\Ga$.
These terms are absent for co-compact group $\Ga$ (compact hyperbolic manifolds).
Furthermore, the contribution of hyperbolic elements is exponentially
small in $t$. Thus, in general, the result of Proposition 2 still holds
true.

Let us analyse the consequences of the presence of logarithmic terms
in the latter expansion.
As usual, we may introduce the
$\ze$-function associated with the elliptic operator $L_\de$
by means of the Mellin transform
\beq
\ze(s|L_\de)=\frac{1}{\Ga(s)}\int_0^\ii dt\:t^{s-1}
\Tr e^{-t L_\de}
\:,\label{zf}\eeq
valid for $\Re s>3/2$. In order to get the meromorphic structure of the
function (\ref{zf}), we
splits the integration range in the two intervals $[0,1)$ and
$[1,\ii)$ obtaining in this way two integrals.
The last one is regular for $s\to0$, while the behaviour of the
first one can be estimated by using the asymptotics, Eq.~(\ref{l3}).
Thus we have
\begin{proposition}
The meromorphic structure of the $\zeta$-function reads
\beq
\ze(s|L_\de)=\frac{1}{\Ga(s)}
\sum_{r=0}^\ii\aq\frac{A_r}{s+r-\frac{3}{2}}
-\frac{P_r}{(s+r-\frac{3}{2})^2}\cq
+\frac{J(s)}{\Ga(s)}\:,\eeq
where $J(s)$ is an analytic function.
\end{proposition}
>From this, it follows that
the analytic continuation of $\zeta$-function
is  regular at $s=0$. It has to be noted
also the presence of  double poles, caused by the logarithmic terms.

We conclude this Section by computing the asymptotic behaviour for very
large $\de$ of the derivative of $\zeta$-function evaluated in zero. To
this aim, again the asymptotic behavior for small $t$
gives \cite{voro87-110-439}
\begin{proposition}
\beq
\ze'(0|L_\de)=\frac{V(F)\de^3}{6\pi}+\frac{1}{2}\de\ln\de
-\de\at C+\frac12\ln2+\frac12\ct +O(1/\de)
\:.\label{asd}\eeq
\end{proposition}

\s{The functional determinant}

In this Section, making use of the trace formula, we shall
compute the functional determinant of a Laplace-type operator on $H^3/\Ga$.
We briefly explain the method which is based on $\zeta$-function
regularization and an evaluation by quadratures with
an appropriate  choice of the function $h(r)$
appearing in the trace formula
\cite{sarn87-110-113,voro87-110-439,efra88-119-443}.
The $\zeta$-function,
for $\Re s$ sufficiently large, can be rewritten in the form
\beq
\ze(s|L_\de)=\sum_\si\rho_\si\at\la_\si+\de^2-1\ct^{-s}
=\sum_i \at \la_i+\de^2-1\ct^{-s}
+\int_0^\ii \at \la+\de^2-1 \ct^{-s}\:\rho_\la\:d\la
\:,\label{zf1}\eeq
where the sum over $i$ run over the discrete
spectrum, $\la_i$ being the eingenvalues. For the continuous spectrum,
$\rho_\la$ is proportional to the
logarithmic derivative of the S-matrix $S(s)$. One has
\beq
\ze'(s|L_\de)=-\sum_\si \rho_\si\at\la_\si+\de^2-1\ct^{-s}
\ln(\la_\si+\de^2-1)
\:.\label{zf11}\eeq
>From the latter equation one gets
\beq
\frac{d}{d \de} \at \frac{1}{2\de}\frac{d}{d \de}\ze'(s|L_\de)
\ct=2\de
\sum_\si \rho_\si  \at \la_\si+\de^2-1 \ct^{-s-2} +O(s)
\:.\label{zf2}\eeq
A standard Tauberian argument and Eq.~(\ref{l3}) gives a Weyl's
estimate for large $\si$, namely
$(\la_\si+\de^2-1)^{-1}\simeq \si^{-2/3}$.
As a consequence, in the limit $s \to 0 $, the r.h.s. of Eq.~(\ref{zf2})
is finite. This works for $D=2$ as well as for $D=3$ dimensions.
In higher dimensions it is necessary
to take further derivatives with respect to $\de$ \cite{voro87-110-439}.

On the other hand, we may  rewrite the formula
Eq.~(\ref{stf}) as  (here $r^2+1=\la$ and $\ga$ is the identity or parabolic
element in $\Ga$)
\beq
G(\de)=\sum_\si\rho_\si h_\de(r_\si)
=\sum_j h_\de(\la_j)-\frac{1}{4\pi}\int_{-\ii}^\ii
h_\de(r)
\frac{S'(1+ir)}{S(1+ir)}\:dr+\frac{h_\de(0)}{4}S(1)
\label{fp}\:.
\eeq
$G(\de)$ denoting the "geometrical" part
\beq
G(\de)=V(F)\int_0^\ii\frac{r^2}{2\pi^2}\:h_\de(r)\:dr
+ Cg(0)+\frac{h_\de(0)}{4}
-\frac{1}{4\pi}\int_{-\ii}^\ii h_\de(r)
\psi(1+\fr{ir}{2})\:dr
\:.\label{stfg}\eeq
Let us choose the function $h_\de$ as
\beq
 h_\de(r)=\frac{1}{r^2+\de^2}-\frac{1}{r^2+a^2}\,,\hs
g_\de(0)=\frac{1}{2\de}-\frac{1}{2a}
\:,\label{h}\eeq
with $a$ a non vanishing constant.
Taking the derivative with respect to $\de$ we have
\beq
 2\de \sum_\si \rho_\si  \at \la_\si+\de^2-1 \ct^{-2} =-\frac{d}{d
\de}G(\de)
\:.\label{stf2}\eeq
Making the comparison between Eqs. ~(\ref{zf2}) and ~(\ref{stf2}),
integrating twice in the variable $\de$ and taking the limit $s\to0$
we finally obtain
\beq
\ze'(0|L_\de)=-2\int\de G(\de)\:d\de+c_1 \de^2+c_2
\:,\label{S}\eeq
where the constants $c_1$ and $c_2$ can be determined from the
asymptotics for large $\de$.
The primitive related to the geometrical part can be easily computed
by making use of the Selberg trace formula with the choice ~(\ref{h}).
One has
\beq
G(\de)=-\frac{V(F)}{4\pi}(\de-a)
+\frac{ C}2\at\frac1\de-\frac1a\ct
+\frac14\at\frac1{\de^2}-\frac1{a^2}\ct
-\frac14\aq\frac{\psi(1+\de/2)}\de-\frac{\psi(1+a/2)}a\cq
\:.\label{bu}
\eeq
As a consequence,
\beq
\ze'(0|L_\de)= \frac{V(F)\de^3}{6\pi}+[c_1+Q(a)]\de^2
- C\de-\frac{1}{2}\ln\de
+\ln\Ga(1+\fr{\de}{2})
+c_2
\:,\label{S1}\eeq
where
\beq
Q(a)=\frac{1}{4a^2}-\frac{V(F)a}{4\pi}+\frac C{2a}
-\frac{\psi(1+a/2)}{4a}
\:.\label{}\eeq
The inclusion of the contribution related to the hyperbolic
elements in Eq.~(\ref{S1}) is almost straightforward and can be found in Refs.
\cite{eliz94b,byts96-266-1}. It is additive and reads simply $\ln Z(1+\de)$,
$Z(s)$ being the Selberg zeta-function.

In the large $\de$ limit, $\ln Z(1+\de)$ is vanishing and one has
\beq
\ze'(0|L_\de)\simeq\frac{V(F)\de^3}{6\pi}
+[c_1+Q(a)]\de^2 +\frac12\de\ln\de
-\de\at C+\frac12\ln2+\frac12\ct
+\frac12\ln\pi+c_2
\:,\label{asd1}\eeq
which agrees with Eq.~(\ref{asd}) if
\beq
c_1=-Q(a)\:,\hs c_2=-\frac{1}{2}\ln\pi
\:.\label{cf}\eeq
Summarizing we have proved the
\begin{theorem}
One has the identity
\beq
\det L_\de=\frac2{\sqrt{\pi\de}\Ga(\frac\de2)}
\exp\at-\fr{V(F)\de^3}{6\pi}+C\de\ct\, Z(1+\de)
\:.\label{detfin}\eeq
\end{theorem}

\s{Conclusions}

In this paper we have computed the functional determinant of a
Laplace-like operator on a non-compact 3-dimensional hyperbolic
manifold with finite volume fundamental domain,
by the method of quadratures. In addition the contributions to the heat kernel
and $\zeta$-function associated with identity and parabolic elements of
isometry group is analysed.
The constant appearing in the
quadrature process has been determined by means of the asymptotic behavior of
the functional determinant, which may be achived again making use of the trace
formula for the heat kernel.
This method is particular useful in the evaluation of the functional
determinants, because it permits to avoid the problem of finding the
analytical continuation of the zeta-function, which may present
computational difficulties. On the other hand, the method requires the
existence of a trace formula and its validity can be extended to
more general cases (see for example
\cite{gang77-21-1,will92b,will95-191-245}).

\ack{ We thank Prof.~F.L.~Williams for useful remarks. A.A.B.~would like
to thank I.N.F.N. for financial support and Prof.~M.~Toller for kind
hospitality at Department of Physics, University of Trento.
The research of A.A.B.~was supported in part
by Russian Foundation for Fundamental Research
grant No.~95-02-03568-a and by
Russian Universities grant No.~95-0-6.4-1.}


\end{document}